\documentclass{elsart}

\usepackage{natbib}

\usepackage{graphics}
\usepackage{graphicx}
\usepackage{epsfig}

\usepackage{amssymb}

\begin{document}

\begin{frontmatter}



\title{The recurrent nova RS Oph: A possible scenario for type Ia supernovae}


%\author{}

%\address{}

\author{M. Hernanz}

\address{Institut de Ci\`encies de l'Espai (CSIC) and
Institut d'Estudis Espacials de Catalunya (IEEC), Campus UAB,
Facultat de Ci\`encies, Torre C5 - parell - 2a planta,
08193 Bellaterra (Barcelona), Spain}
\author{J. Jos\'e}
\address{Departament de F\'{\i}sica i Enginyeria Nuclear, EUETIB 
(UPC), C/Comte d'Urgell 187, 08036 Barcelona, Spain, and IEEC}

\begin{abstract}
The recurrent nova RS Oph experienced an outburst in 2006, 21 years
after its previous explosion in 1985, as expected. It was observed
at almost all wavelengths, and important information about its properties
is still being extracted. We present theoretical models of the explosion
of this fascinating object, which indicate that the mass of the accreting
white dwarf should be very close to the Chandrasekhar mass, to allow for
such a short recurrence period. In addition, since models suggest that 
this nova ejects less 
mass than it accretes, it is an excellent candidate for a thermonuclear
supernova explosion, in about $10^5-10^7$ years from now. We also analyze
the emission of soft gamma-rays by RS Oph detected with the BAT instrument
onboard Swift, and with the PCA onboard RXTE. We rule out that this emission 
has its origin in radioactive decays in the expanding nova envelope.
\end{abstract}


\end{frontmatter}

% main text
\section{Introduction}
\label{intro}

Recurrent novae are defined as systems with more than one recorded nova outburst. 
There are few members of the recurrent novae 
subclass: just 10 (see \citet{AM99} and review by \citet{Miko08}), 
divided in two groups according to their orbital periods. RS Oph belongs to the 
long period group, with P$_{\rm orb}=456$ days \citep{Fekel00}, 
which are interpreted as interacting 
binaries  where a hot white dwarf accretes mass from a giant secondary; these 
recurrent novae are a subclass of symbiotic binaries. In contrast, classical novae 
occur in cataclysmic variables, a different type of binary system where a white 
dwarf accretes mass from a main sequence companion, overflowing its Roche lobe; typical 
periods of cataclysmic variables are about a few hours.

RS Oph has undergone various recorded outbursts (1898, 1933, 1958, 1967, 
1985 and 2006, and two additional possible outbursts in 1907 and 1945). 
In the two last outbursts, a lot of observational data was gathered. 
The recent outburst in 2006 offered the opportunity to observe RS Oph at 
practically all wavelengths, radio, IR, optical, X-rays 
and soft gamma-rays. However, the very important UV range was missing, since 
there was not a satellite ``equivalent'' to IUE flying in 2006. It was a pity, since 
IUE made very important observations in the previous RS Oph eruption in 1985 
\citep{Shore96}, which could not be repeated in 2006. 
On the other hand, current space missions like Spitzer, RXTE, XMM-Newton, Chandra 
and Swift have provided a wealth of IR and X-rays data, of much better quality 
than what could be 
obtained at such energy ranges in the previous outburst. All in all, there is now 
an impressive data set; its interpretation  is a real challenge for theorists 
and it has provided already many exciting results. 

It is worth noting that there is not yet a self consistent model of the RS Oph 
outburst itself, since it is not 
straightforward to find a set of initial conditions leading to a thermonuclear 
runaway (TNR) with a recurrence period as short as 21 years; classical novae recurrence 
periods are much longer ($\ge 10^4-10^5$ years), well in agreement with the expected 
time to reach outburst conditions with typical accretion rates and initial white dwarf 
masses in cataclysmic variables. The only known way to reconcile a short 
interoutburst period -about some decades- with a TNR (i.e., 
explosive hydrogen burning in degenerate conditions on top of the white dwarf), 
is that a very massive white dwarf, close to the Chandrasekhar limit, accretes with a 
large mass accretion rate. This fact gives an additional interest to RS Oph and 
its relatives, since they are potential 
progenitors of thermonuclear -or type Ia- supernovae, in a relatively short time, 
if the white dwarf grows in mass in each nova eruption. Our evolutionary models indicate 
that ejected mass is smaller than accreted mass, so confirming that possibility.  
   
In this paper we present our recent attempts to model the basic properties of the RS 
Oph recurrent nova outburst, with special emphasis on the ratio of ejected to 
accreted mass, the ejecta chemical composition and the potential gamma-ray emission. 
The role played 
by the initial conditions of the white dwarf is underlined. We also discuss the 
impossibility to explain the Swift/BAT hard X-soft gamma-rays detections as related 
with the continuum gamma-ray emission coming from electron-positron annihilation in 
the expanding ejecta. 

\section{Main observational properties}
\label{scenario}

The last eruption of RS Oph was recorded on 2006 February 12.83 UT, reaching magnitude 4.4 
\citep{Narumi06}. Observations at other wavelengths have provided important data. For 
instance, radio observations yielded a new detection (not obtained in the previous 
outburst) of early emission at low radio frequencies, clearly of 
nonthermal origin  
\citep{Kantharia07}, and high 
resolution VLBA images, clearly showing an asymmetric expanding shock wave 
\citep{Obrien06}. Early X-ray 
observations with RXTE \citep{Sokoloski06} and Swift/BAT \citep{Bode06} showed hard X-ray 
emission (up to 50 keV), a clear indicator of the blast wave from RS Oph, i.e. an outward 
propagating shock 
wave consequence of the interaction between the expanding ejecta and the 
red giant wind (the latter much 
denser than typical circumstellar material in classical novae). Additional data about 
the early 
evolution came from IR observations, which also showed the temporal evolution of the 
shock 
velocity \citep{Das06,Evans07}. There is a difference in the shock velocities deduced 
from the 
X-ray and IR data, which has been shown to indicate efficient particle acceleration; 
nonlinear diffusive shock acceleration of cosmic rays also explains the 
decceleration of the blast wave in RS Oph, which occured faster than predicted by the 
standard adiabatic shock wave model \citep{TH07}. IR observations also provide 
an estimate of the ejected mass: $\sim 3 \times 10^{-6} {\rm M}_{\odot}$ \citep{Das06}.
Another important set of observations in X-rays 
(with Swift/XRT, XMM-Newton and Chandra) revealed highly variable soft X-ray emission at 
about 30 days post outburst, and lasting for about 60 days, with high temporal variability 
\citep{Bode06,Ness07,Nelson08}. The short duration of the supersoft X-ray phase 
points again to a very massive white dwarf in RS Oph \citep{HKL07}. This supersoft 
X-ray emission results from residual hydrogen burning on top of the white dwarf.

It is important to know that overabundances of lithium have been found in RS Oph in 
quiescence, and also in another recurrent nova, T CrB, sharing its basic properties with RS Oph 
\citep{Wallerstein08}. 

\section{Models}
\label{models}

The model for classical novae explosions is based on the thermonuclear runaway
ensuing hydrogen 
degenerate burning, on top of an accreting white dwarf, which drives mass ejection at 
large velocities (hundreds to thousands of km/s). For typical initial masses and 
luminosities of the white dwarf, and mass accretion rates from the secondaries, 
there is an explosion every $10^4-10^5$ years. 
Some kind of mixing between the accreted matter and the underlying white 
dwarf should be invoked to explain the large enhancements above solar values 
of CNO nuclei, and also of neon, found in the ejecta of many novae (a list of 
abundances in novae ejecta can be found  in \citet{Gehrz98}). 
However, no large overabundances 
of metals were found in RS Oph \citep{Shore96}, so that not much mixing is expected to 
occur. The main difficulty with the ``standard''  TNR scenario 
for RS Oph is the short recurrence period, ${\rm P_{rec}}= 21$ years. Both short 
${\rm P_{rec}}$ and lack of overabundances above solar in the ejecta
point to fast evolution, only attainable with some particular 
combinations of large white dwarf mass and accretion rate. A first set of nova models 
on white dwarfs with M=1.38 ${\rm M}_{\odot}$ were computed by \citet{Starr85}. 
At nearly the 
same epoch, an alternative scenario for RS Oph outburst was invoked \citep{Livio86}, 
casting doubts on the TNR model and assuming a main sequence companion instead of a giant. 
However, it is nowadays accepted that the companion of the white dwarf in RS Oph is a 
red giant and that a TNR is responsible for its outburst. One observational fact 
that proves this scenario is the residual steady H-burning revealed 
by the supersoft X-ray emission detected $\sim 30$ days after outburst. Other arguments  
come from the shock wave revealed both in X-rays, IR and radio. 

We reanalyse here the TNR model for RS Oph, by means of the hydrodynamic code described 
in \citet{JH98}, which follows the evolution of the accreting white dwarf from the 
accretion phase up to mass ejection, with complete nucleosynthesis  included, 
and updated nuclear reaction rates. 
The initial conditions are very relevant for the ensuing properties of the explosion. 
An initial mass always larger than 1.2 ${\rm M}_{\odot}$ has been adopted, with a range 
of mass accretion rates and white dwarf luminosities. No mixing 
of any kind between the accreted mass (assumed to have solar composition) 
and the white dwarf core has been adopted. 
The main issue is the initial white dwarf luminosity; according 
to \citet{AM99}, white dwarfs in symbiotic stars are quite luminuous, as a consequence of 
steady H-burning on their surface. However, such high luminosities do not mean that the 
whole white dwarf core is very hot, but just its outer envelope. Therefore, once the 
purely symbiotic phase ends, a central core temperature-luminosity relation is again valid, 
and relatively low luminosities, such as those required to power a TNR, are possible. 

A search for appropriate initial conditions leading to short recurrence periods is the 
first step to model RS Oph. The recurrence period can be written as 
${\rm P_{rec}}= \Delta {\rm M_{acc}/\dot M} =21$ years, where $\Delta {\rm M_{acc}}$ 
is the accreted mass on top of the white dwarf required to power the outburst through 
a TNR and $\dot {\rm M}$ is the mass accretion rate. The question is which combinations 
of the initial white dwarf mass, ${\rm M_{wd}^{ini}}$, 
$\dot {\rm M}$ and the initial luminosity, ${\rm L_{wd}^{ini}}$, 
lead to a nova outburst with such a short period of 21 years. We have adopted a range 
of ${\rm L_{wd}^{ini}}$ between $10^{-2}$ and 1 ${\rm L_\odot}$, and two values for 
$\dot {\rm M}$ ($2 \times 10^{-7}$ and $10^{-8} {\rm M_\odot/year}$). The initial 
white dwarf mass ranges from 1.25 to 1.38 ${\rm M}_\odot$. The chemical composition 
of the white dwarf should be ONe (and not CO) according to standard stellar evolution 
(see for instance \citet{Rit96}). 

% FIGURE

\begin{figure}
%\label{ignition}
\centering
\psfig{file=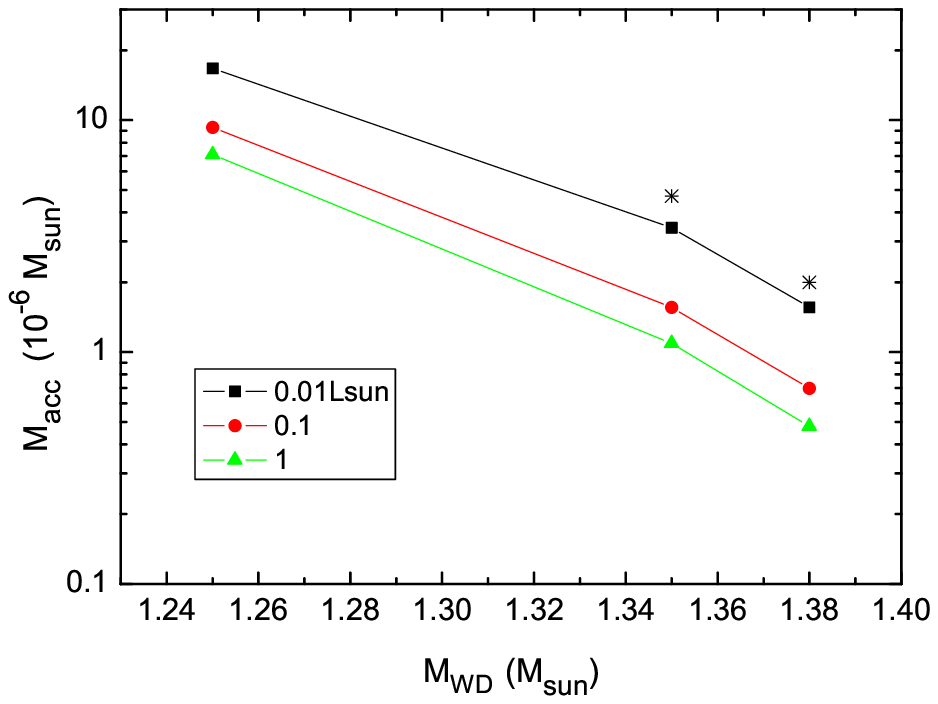}
\psfig{file=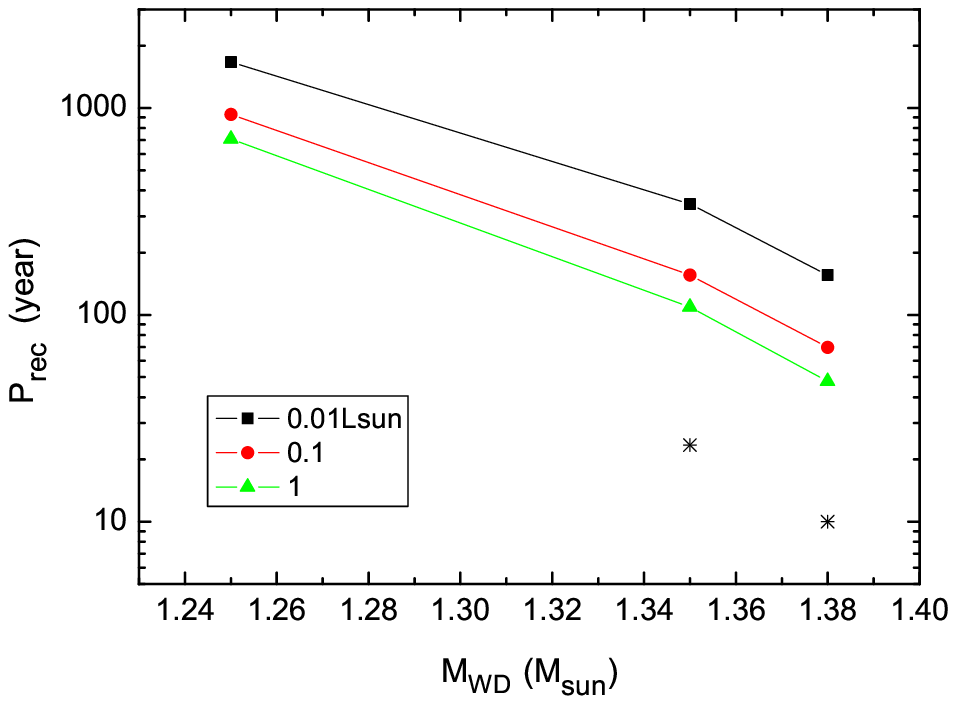}
\caption{Upper panel: accreted masses required to reach hydrogen ignition conditions and 
to power a TNR. Lower panel: recurrence periods for the same 
set of initial conditions as in the upper panel. Asterisks denote the values obtained for 
$\dot {\rm M}= 2 \times 10^{-7} {\rm M_\odot/year}$ and 
${\rm L_{wd}^{ini}}= 10^{-2} {\rm L_\odot}$.}
\end{figure}

We show in Figure 1 (upper panel) the accreted masses needed to reach 
hydrogen ignition conditions versus the white dwarf initial mass, for various luminosities 
and $\dot {\rm M}= 10^{-8} {\rm M_\odot/year}$. We also include the results for  
$\dot {\rm M}=2 \times 10^{-7} {\rm M_\odot/year}$ and
${\rm L_{wd}^{ini}}=10^{-2} {\rm L_\odot}$. The corresponding recurrence periods 
are shown in the lower panel of the same figure.
One sees at first glance that the critical accreted mass does not 
depend only on ${\rm M_{wd}^{ini}}$, as some basic approximate formula indicate (since  
they only require that some pressure at the base of the accreted envelope -about 
$10^{21-22}$ dyn/cm$^2$- is reached to power degenerate H-ignition). As expected,  
an increase in the 
initial luminosity reduces the accreted mass needed to power the TNR and thus the 
recurrence period. 
Another result is that the decrease in accreted mass and recurrence period is steeper as 
the initial white dwarf mass approaches the Chandrasekhar limit, since the star becomes 
easier to destabilize. The main issue is that recurrence periods as short as 21 years are 
really hard to obtain; only masses larger than $\sim 1.35 {\rm M}_\odot$ and accretion rates 
about $10^{-7}{\rm M_\odot/year}$ can provide them. It is worth noting that the initial 
luminosity should be reasonably low, ${\rm L_{wd}^{ini}}= 10^{-2} {\rm L_\odot}$, 
to guarantee a TNR. Another possibility is a smaller accretion rate, 
$10^{-8}{\rm M_\odot/year}$,  with larger L, 1 ${\rm L_\odot}$, but the explosion is then 
less powerful and marginally compatible with observations. 
 
The high accretion rates required agree with some recent predictions from 3D 
simulations, yielding a red giant mass loss rate of about $10^{-7} {\rm M_\odot/year}$ 
and an accretion rate of about 10\% this value, i.e. $10^{-8} {\rm M_\odot/year}$ 
\citep{WFS08}. It is worth mentioning, however, that observations by 
\citet{Wor07} have detected the resumption of optical flickering in RS Oph, 
indicating that mass accretion had been restablished, with deduced 
accretion rates spanning the range between $10^{-10}$ and  
$10^{-9} {\rm M_\odot/year}$, depending on the mechanism of accretion (either 
wind accretion or Roche lobe overflow); with such small accretion rates, recurrence 
periods would be larger than 21 years by orders of magnitude \citep{JH98}.

In Table 1 we show some properties of the explosion for the most interesting cases.  
The models with  
${\rm M_{wd}^{ini}}$=1.35 and 1.38 ${\rm M}_\odot$, 
$\dot {\rm M}= 2 \times 10^{-7} {\rm M_\odot/year}$, 
${\rm L_{wd}^{ini}}= 10^{-2} {\rm L_\odot}$ illustrate 
the effect of the initial white dwarf mass: a slight increase in initial mass 
from 1.35 to 1.38 ${\rm M}_\odot$ reduces the recurrence period by more than a factor 
of 2 (from 24 to 10 years). The properties of two additional models, with  
$\dot {\rm M}= 10^{-8} {\rm M_\odot/year}$ and two initial luminosities - 
${\rm L_{wd}^{ini}}= 10^{-2}$ and 1 ${\rm L_\odot}$ - are also displayed, to 
illustrate the effect of the initial white dwarf luminosity and of the mass 
accretion rate. A decrease in the accretion rate by a factor of 20 increases the 
recurrence period by a similar factor (from $\sim 10$ to $\sim 160$ years), 
provided that the 
initial luminosity is unchanged ($10^{-2}{\rm L_\odot}$). On the other hand, an 
increase by a factor of 100 in the initial luminosity (from $10^{-2}$ to 
1 ${\rm L_\odot}$) just reduces the recurrence period by a factor of 2.7 (from 
$\sim 160$ to $\sim 60$ years), provided that the accretion rate and the initial 
white dwarf mass remain unchanged (1.38 ${\rm M}_\odot$ and 
$10^{-8} {\rm M_\odot/year}$, respectively).

Peak temperatures (see again Table 1) range around $(2-3) \times 10^8$ K, which 
are moderate but similar to those of classical novae; 
in contrast, the total time elapsed since accretion starts up to ignition is much 
shorter, as already emphasized. The corresponding accreted and ejected masses are 
quite small, 
$\sim 10^{-6}$ M$_\odot$ or even smaller ($\sim 10^{-7}$ M$_\odot$ for the model with 
large initial luminosity, 1 ${\rm L_\odot}$). However, the case 
with large initial luminosity does not give a very energetic explosion and would not 
well represent RS Oph (but it is not too bad concerning the recurrence period, 
58 years, just a factor of $\sim 3$ above the observed value). The two models with 
$\dot {\rm M}= 2 \times 10^{-7} {\rm M_\odot/year}$ are thus those that best reproduce 
RS Oph main observed features. The maximum bolometric luminosities reach 
$\sim (3-5) \times 10^{5}$L$_\odot$, the average global 
kinetic energy is about $10^{44}$ erg, with average velocities of 
$(3-4) \times 10^{3}$ km/s, and ejected masses are $\sim 10^{-6}$ M$_\odot$, 
which indicate that this nova is a bit less 
energetic than classical ones, and with a smaller content of radioactive 
nuclei (as shown below). 

% TABLE

% Table
\begin{table}
%\vspace{-2.0cm}
\caption{Main properties of the RS Oph models: peak temperature, accreted 
and ejected masses, timescales (accretion, recurrence time),  
total white dwarf mass increase and time needed to reach ${\rm M_{Chandra}}$. 
Units are: $10^8$ K for T, M$_\odot$ for masses, years for times.}
\smallskip
\begin{center}
%{\small
\begin{tabular}{cccccccc}
\hline\hline
M$_{\rm WD}$ & T$_{\rm peak}$ & M$_{\rm acc}$   & M$_{\rm ej}$ 
             & t$_{\rm acc}$  & t$_{\rm rec}$   & $\Delta {\rm M_{wd}}$
             & $\Delta {\rm t_{Chandra}}$\\
\hline\hline
\multicolumn{8}{c}{L$_{\rm ini}=10^{-2}$ L$_\odot$ ; $\dot {\rm M}= 2 \times 10^{-7} 
{\rm M_\odot/year}$}\\
\hline
1.35         & 2.8            & 4.7E-6         & 3.0E-6
             & 23.3           & 24             & 1.7E-6               
             & 6.9E5\\
1.38         & 3.1            & 2.0E-6         & 1.3E-6
             & 10.0           & 10.4           & 0.7E-6
             & 2.9E5 \\
\hline
\multicolumn{8}{c}{L$_{\rm ini}=10^{-2}$ L$_\odot$ ; $\dot {\rm M}= 10^{-8} 
{\rm M_\odot/year}$}\\
\hline
1.38         & 3.0            & 1.6E-6         & 1.1E-6 
             & 156            & 157            & 4.1E-7         
             & 7.6E6 \\
\hline
\multicolumn{8}{c}{L$_{\rm ini}=1$ L$_\odot$ ; $\dot {\rm M}= 10^{-8} {\rm M_\odot/year}$}\\
\hline
1.38         & 2.5            & 4.8E-7         & 4.0E-7 
             & 47.8           & 57.6           & 7.7E-8               
             & 1.2E7  \\
\hline\hline
\end{tabular}
%}
\end{center}
%\vspace{-1cm}
\end{table}

An important result is that the accreted masses are larger than the ejected ones, 
so that the mass of the white dwarf increases: M$_{\rm acc}$- M$_{\rm ejec}$ 
ranges between 0.7 and 1.7 $\times 10^{-6}$ M$_\odot$, leading to an 
expected time to reach the Chandrasekhar mass (M$_{\rm Chandra}$= 1.4 M$_\odot$) 
ranging between 3 and 7 $\times 10^5$ 
years for the two models with $\dot {\rm M}= 2 \times 10^{-7} {\rm M_\odot/year}$.
The other models would need much larger times to reach the Chandrasekhar mass, and thus 
would not be viable type Ia progenitors.

The nucleosynthesis in our models provides large hydrogen and helium abundances in 
the ejecta (mass fractions $\sim 0.6$ and $\sim 0.4$, respectively). There is not 
large overproduction with respect to solar of metals, in agreement with observations, 
and contrary to what is 
obtained in general for classical novae. The main reason for such a difference
is the lack of initial mixing and the rapid evolution to the runaway. Regarding 
$^7$Li, it is not at all overproduced. 
The key issue in the synthesis of $^7$Li is the amount
of $^7$Be (which is transformed into $^7$Li in the ejecta
after an electron capture) that survives the TNR (see \citet{Her96,JH98}).
At the early stages of the TNR, $^7$Be is efficiently destroyed
by proton capture reactions, but as soon as enough T is
reached (about $(5-7)\times 10^8$K) photodisintegration
of $^8$B drives a pseudo-equilibrium between 
$^7$Be and $^8$B. Hence, the final amount of $^7$Li strongly 
depends on the level of $^7$Be destruction prior to this 
pseudo-equilibrium. The faster the TNR, the shorter to reach the 
pseudo-equilibrium stage, and hence, the larger the final amount 
of $^7$Li. This explains why CO models lead in general to larger amounts
of $^7$Li in the ejecta as compared with ONe models (since mixing
with the underlying CO WD brings $^{12}$C fuel into the envelope
which speeds up the TNR dramatically). Also, the increase
of strength in explosions hosting massive white dwarfs explains
why very massive ONe white dwarfs (close to the Chandrasekhar mass)
eject larger amounts of $^7$Li than not so massive ONe white dwarfs, but never 
reaching solar values or larger. Thus, 
the Li overabundance found in RS Oph in quiescence \citep{Wallerstein08} is likely 
not synthesized during the nova outburst. 

There are small amounts of the radioactive nuclei relevant for $\gamma$-ray emission 
($^{13}$N, $^{18}$F, $^{7}$Be and $^{22}$Na, see \cite{Gom98}),   
i.e., much smaller yields in mass fraction (and also absolute yields, because 
of the small ejected masses) than for classical novae.  In addition, the envelope becomes 
transparent to $\gamma$-rays later than in normal novae, because of the 
smaller global kinetic energy. All in all, the expected gamma-ray fluxes are very 
small. This rules out the interpretation of the observed soft $\gamma$-rays 
from RS Oph, with the BAT instrument on board the Swift satellite 
\citep{Bode06}, as due to the decay of radioactive nuclei. Shocks within the 
ejecta should provide a better explanation of the observed hard X-ray emission 
\citep{Sokoloski06,Bode06}. In fact, one could know beforehand that 
Swift/BAT observations of RS Oph can not be 
interpreted as due to radioactive nuclei decay, by two 
reasons: the spectral shape and the light curve 
\citep{Gom98,HGJ02}. The shapes of the $\gamma$-ray spectra expected from radioactive 
decay at various 
epochs are such that there is less emission in the 15-25 keV energy band than 
in the 25-50 keV band, due to the sharp spectral cutoff at low energies 
produced by photoelectric absorption. The contrary 
was observed with Swift/BAT and RXTE, as shown in Figure 1 from 
\cite{Bode06}. Concerning the $\gamma$-ray light curve, our models indicate 
that the maximum in visual flux occurs later than the maximum in 
$\gamma$-rays (simultaneous to peak temperature), for all types of novae, 
again in contradiction with the observations with Swift/BAT, where the peak 
in hard X/soft $\gamma$-ray emission 
occurred after the optical outburst (see Figure 1 from \cite{Bode06}).

\vspace{-0.6cm}
\section{Discussion and conclusions}
\label{conclu}
Models of the RS Oph explosion should account for its main observed properties, e.g., 
very short recurrence period (21 years), small ejected mass, roughly solar chemical 
composition of the ejecta and maximum luminosities and ejecta velocities similar to 
those of classical novae. Hydrodynamic models of massive (larger than 
$\sim$ 1.35 M$_\odot$) accreting white dwarfs with 
large accretion rates - ranging from  $\dot {\rm M}= 10^{-8}$ 
to $2 \times 10^{-7} {\rm M_\odot/year}$ - well reproduce these features. But there are 
still some aspects to be better understood, as the need of initially low 
white dwarf luminosities, as compared with some observations at quiescence. 
We want to point out a few important puzzles.
First of all, very massive white dwarfs should be made of ONe (if they were 
born as massive white dwarfs); therefore, they pose a problem as scenarios for 
type Ia supernovae, because ONe white dwarfs rather collapse to a neutron star 
than explode, when reaching M$_{\rm Chandra}$ (see \citet{GCGB05} and 
references therein). One could envision that the white dwarf was 
initially born as a CO white dwarf, but then it should have been less massive 
than 1.1 M$_\odot$ and be able to reach 1.35-1.38 M$_\odot$ through 
successive recurrent nova eruptions. However, this seems unrealistic because 
models do not reproduce such behaviour.
Second, the mass left after the 
recurrent nova explosion, i.e., M$_{\rm acc}$- M$_{\rm ejec}$ (see above) 
should be compatible with the duration of the super soft X-ray emission 
phase (around 60 days, see \cite{HKL07}). According to our models 
\citep{SH05}, the 
remnant envelope mass for a hydrogen mass fraction of 0.6 and mass of the 
white dwarf larger than 1.3 M$_\odot$ should be around 
$10^{-7}$ M$_\odot$, and surely smaller than $10^{-6}$ M$_\odot$. 
Therefore, the mass of the white dwarf is more plausibly 1.38 than 1.35 
M$_\odot$ (see Table 1).
Finally, the hard X/soft $\gamma$-rays detected with Swift/BAT and RXTE in the early 
phases of the explosion could not come from electron-positron annihilation emission 
(with positrons coming from the decay of the $\beta^+$-unstable nuclei 
$^{13}$N, $^{18}$F in the ejecta), both because of the tiny amount of radioactive 
nuclei, the too late appearence of such emission and its spectral shape. Deeper 
studies of both the thermonuclear runaway phase and the interaction between the 
ejecta and the dense stellar wind of the red giant are in progress, 
to better understand the 
wealth of observational data available for the fascinating RS Oph recurrent nova. 

\section*{Acknowledgements}
This work has been supported by the spanish MEC 
and by the catalan DURSI through grants ESP2007-61593, AYA2007-66256 and 
2005SGR00378, and by FEDER funds. We thank Vincent Tatischeff for many 
valuable discussions.

%\newpage

\end{document}